# An Intrinsically Knowledge-Transferring Developmental Spiking Neural Network for Tactile Classification

Jiaqi Xing, Libo Chen, ZeZheng Zhang, Mohammed Nazibul Hasan, Zhi-Bin Zhang*

Solid-State Electronics, Department of Electrical Engineering, Uppsala University, Uppsala, 75121, Sweden

*Correspondence to: zhibin.zhang@angstrom.uu.se

**Abstract:**

Gradient descent computed by backpropagation (BP) is a widely used learning method for training artificial neural networks but has several limitations: it is computationally demanding, requires frequent manual tuning of the network architecture, and is prone to catastrophic forgetting when learning incrementally. To address these issues, we introduce a brain-mimetic developmental spiking neural network (BDNN) that mimics the postnatal development of neural circuits. We validate its performance through a neuromorphic tactile system capable of learning to recognize objects through grasping. Unlike traditional BP-based methods, BDNN exhibits strong knowledge transfer, supporting efficient incremental learning of new tactile information. It requires no hyperparameter tuning and dynamically adapts to incoming data. Moreover, compared to the BP-based counterpart, it achieves classification accuracy on par with BP while learning over ten times faster in ideal conditions and up to two or three orders of magnitude faster in practical settings. These features make BDNN well-suited for fast data processing on edge devices.

**Main Text:**

The rapid and extensive adoption of deep learning through real-valued artificial neural networks (ANNs) has significantly influenced the world, although at a remarkably high energy cost (*1–5*). In contrast, the utilization of spiking neural networks (SNNs), as a bio-plausible and new generation of ANNs, presents a promising alternative characterized by lower power consumption and higher computational efficiency (*3, 6–8*). However, unlike real-valued ANNs that have been widely trained by using gradient descent algorithm, with gradients computed through backpropagation (BP) process, SNNs often grapple with training difficulties caused by the non-differentiable of spikes. It has been recently demonstrated that the BP-based learning procedure can be applied to train SNN through ANN-to-SNN conversion (*9–13*) or surrogate function methods (*14–16*) although the resulting performance can be lower as compared to the real-valued ANN counterpart.

Although the BP-based learning procedure is powerful in training real-valued ANNs and SNNs, it is power-hungry and computationally expensive particularly upon large volume of datasets. It often requires experienced engineer in architecture modulation to, e.g., determine number of neurons. Moreover, due to its catastrophic forgetting, it is generally rather rigid upon new data, for which re-training from scratch is required. Sensors and other edge devices in operation generate frequently new data and rapid data processing is demanded (*17–19*), such rigidity causes the BP-based learning approach incompetent to sensors and edge devices with time-, resource- and power-constraints in real-world, and dynamic environments.

An illustrative application scenario involves tactile perception, a crucial aspect for robots and prosthetic hands to enable sensitive detection of a physical touch to guarantee safety during direct interactions with humans. A second aspect is to reproduce adept manipulation of objects, particularly delicate items akin to human capability (*20–23*). In the quest to develop artificial tactile systems facilitating tactile perception, the prevailing approach entails the generation of analog tactile signals in response to tactile stimulation. This is achieved through the use of electronic (e-) skins, followed by the processing of these signals utilizing real-valued ANNs (*21*, *24–26*). Recently, we have demonstrated a neuromorphic tactile system capable of rapid object classification (*27*, *28*). The system, which is represented as a tactile SNN, was trained by using the BP method. As a consequence, the aforementioned limitations intrinsic to this conventional learning method persists.

To address the challenges associated with the BP-based learning procedure, we present brain-mimetic developmental spiking neural networks (BDNNs), as a fundamentally new learning approach. BDNNs emulate postnatal brain development mainly in two aspects. The first aspect involves growth and pruning of neurons and synapses during formation of a neural circuit driven by neuronal activities associated with experiences. The second aspect pertains to continuing development of neural circuits established with previous experiences. By utilizing tactile signals in the form of spike trains as input data generated by using a neuromorphic e-skin, we demonstrated that BDNN-based learning outperforms the BP-based methods(*29*, *30*), showcasing its comparable classification accuracies, more than 10-time faster learning speed, superior adaptivity without the need of hyperparameter modulation, generalization and capability of knowledge transfer. This learning paradigm with BDNN is anticipated to pave the way for faster, efficient, autonomous, and dynamically adaptive learning, supporting edge computing in close proximity to sensors with power and resource constraints.

**RESULTS**

**Learning with BDNN emulates development of neural circuits**

The evolution of human brain for six million years has created the most energy-efficient and powerful learning methodology used by human. While the majority of primitive neurons have been formed in the early brain development before birth (*31*), the neural circuits undergo continual modification through molecular mechanisms intricately linked to neuronal activities elicited by experiences in postnatal life. The development of a neural circuit involves simultaneous expansion of dendritic and axonal branches, coupled with establishment and pruning of synaptic connections. When human learns objects by grasp, the grasp activates mechanoreceptors in the skin of a hand, generating spike trains that encode the information about the grasp. The information is further processed and transmitted to the brain. Subsequent complex cognitive process results in a network consisting of neurons in different functional areas in the brain. The network represents the concept of the objects learned by grasp.

Imagine that a child after birth plays a set of objects for the first time, the brain initially has no neural circuit associated with these toys in the brain. During grasping a thin cylindric tube that activates mechanoreceptors in the hand (Fig. 1A), a neural circuit in the brain develops by recruiting gradually more neurons into the circuit through neuronal growth, synapse formation and pruning (Fig 1.B). The activated neurons grow in their axonal structure to facilitate pathfinding for signal transmission and, as a result, construct multiple synaptic connections (*32*). Synapse establishment and pruning jointly define the neural circuit $n$ that represents the concept of the thin cylindric tube (Fig 1. C). It has been well established that establishment and pruning of synapses follow the Hebb's learning rule which is governed by the spike-timing dependent plasticity mechanism (*33*).

Moreover, human learns new things much faster and easier with increased previous experience. Human is capable of utilizing previously-learned knowledge in learning incremental new things, unlike the BP-based learning which suffers from catastrophic forgetting. As illustrated in Fig. 1D, the child is provided with additional new objects (e.g., apple, cubic, prismatic cup). When the cylindric tube is grasped, the already-established neural circuit *n* is rapidly activated. During alternative grasping all the objects, the neural circuit *n* is used and incorporated into the expanded neural circuit (Fig 1E) (*32*, *34*). When the development process is completed, the resulting neural circuit *N* represents the information about the four objects (Fig. 1F).

To emulate the human's learning by grasp, our BDNN model (Fig. S1) uses comprehensively spikes for information representation and signal processing. Each neuron is governed by the leaky integrate and fire (LIF) model. Aligning with the principles of developmental neuroscience, BDNN undergoes growth and refinement, driven by sensor input, for object classification. The number of input neurons in a network is 64, which is the number of neuromorphic tactile sensors in our e-skin (SM, Fig. 1G). The tactile sensors, when getting contact with an object, generated time-dependent multiple-channel spike trains. The number of output neurons in the network varies according to the number of objects to be classified. During a learning process with BDNN, incremental hidden neurons are recruited into the network, which prompts pathfinding and pruning in the networks (Fig. 1H, I).

For a fresh learning where a BDNN is constructed from scratch (Fig. 1H), a dataset containing three different objects (cubic, trapezoid, cone) are fed into the network during the increase of the hidden neurons. During the growth, pathfinding and pruning are steered by operations of non-divergence, convergence, bound-tightening, and optimization as elucidated by our mathematical description (SM). As a result, the optimal structure of the current network that maximizes classification accuracy is generated. Iterative addition of hidden neurons results in diverse network architectures, each characterized by variation in classification accuracy. During the addition of new neurons, the pre-formed synapses between the input neurons and the already existing hidden neurons in the network remain unchanged. When the highest classification accuracy is reached, the addition of hidden neuron stops leading to the best BDNN upon the dataset of the three objects. For the simplicity, the resulting BDNN is termed as fresh (*f*-) BDNN(*X*) where *X* is the dataset associated with grasping a specific class of objects.

To emulate the experienced learning of human, a fresh BDNN is initially constructed for representing the cubic upon the dataset of the cubic (Fig 1 I). Subsequently, a larger dataset of the cubic and the trapezoid is fed into the BDNN without adding any new hidden neuron, but increasing the number of output neurons to accommodate the concept of cubic and the trapezoid. The resultant BDNN is used as the seed, and continuously grows until the network classifies the cubic and the trapezoid with a desired accuracy. The resultant BDNN is termed as experienced (*e*-) BDNN for the cubic and the trapezoid. In a similar manner, the *e*-BDNN for the cubic, and the trapezoid can become a seed to further grow to learn a set of objects with additional new object, i.e., the cubic, the trapezoid and the cone.

**Learning with BDNN is adaptive to tactile sensory input**

Datasets generated by grasping objects differ from each other in the content (e.g., features) about objects and data complexities. The difference may depend on the number of objects, types of objects, and also behavioral gestures in gasp actions. Types of objects are characterized by, for example, shapes, stiffness, and roughness. To evaluate the learning performance of BDNN, we start with the fresh learning upon five different 5-class datasets, i.e., *f*-BDNN. A fresh learning refers to develop a *f*-BDNN from scratch driven by sensory input. Each 5-class dataset comprises tactile data in the form of spike trains obtained by grasping of five objects selected from a pool of total 20 objects (Fig S2). Displayed in Fig. 2A is the progression in the training and testing accuracies of a *f*-BDNN upon a 5-class dataset of an apple, a plastic bottle, a solder cleaner, a can box, and a long can (denoted as Class 5-0). During the learning process, the accuracies increase, accompanied by the decrease of the error, with the number of the hidden neurons which are incrementally recruited into the network. Initially, both the training and test accuracies increase rapidly, reaching 90% with 28 recruited hidden neurons. The accuracies continue to rise, although at a slower pace, ultimately reaching 97%. The *f*-BDNN has the nearly identical testing accuracy as its training accuracy, showcasing the high generalization of BDNN (*35*).

When other different 5-class datasets, each of which is associated with different types of objects selected from the pool (SM), were used separately as input, four more *f*-BDNNs were obtained. All the five *f*-BDNN differ in the number of hidden neurons required to achieve similarly high classification accuracies (≥96%) (Fig. 2B). Although each dataset corresponds to the same number of objects, their data complexities vary substantially, as visualized by using the t-distributed stochastic neighbor embedding (t-SNE) method (Fig S3). The complexity of a dataset depends on the number and also the types of objects. Therefore, different complexities of the datasets, even though the datasets contain the same number of objects, give rise to a large structural diversity of the resultant networks (Fig. 2B). It is noticed that the dataset with the largest complexity (Class 5-4, Fig S3E) led to the greatest number of hidden neurons in the network (Fig. 2B). This observation suggests that a BDNN recruits more hidden neurons to represent tactile information with a higher complexity. On the other hand, when a class contains an increased number of objects selected from the pool, both the number of features and the data complexity generally increase, as the t-SNE that visualizes the 5-, 10- and 15-class datasets (Fig. 2C to E). However, as the data complexity depends on types of objects in a class, it can occur that a 10-class dataset (Fig 2F) is more complex than a 15-class dataset (Fig. 2E).

Our results shown in Fig 2B demonstrate that the fresh learning with BDNN is adaptive to large variations caused by different types of objects among the different 5-class datasets. The superior adaptivity of BDNN is further corroborated by fresh learning increased different types of objects (Fig. 2G, H, I). Here, statistical results (Table S1) were presented with several interesting observations. Firstly, all the constructed *f*-BDNNs have ultimately achieved accuracies higher than 90% (Fig. 2G). Secondly, when the number of objects within a class increases, the average accuracy decreases. Meanwhile, the average number of hidden neurons in the BDNN to reach the highest accuracy is increased (Fig. 2H). As a consequence of the increased number of hidden neurons, the space complexity increases with the number of objects (Fig. 2I). Thirdly, as shown in Fig 2H, the distribution in the number of hidden neurons overlap to some extent, particularly for the datasets of 10 and 15 objects. The overlap shows that certain BDNNs for datasets of 15 objects need smaller number of hidden neurons than some for 10 objects. As revealed in Fig 2G, a dataset with a higher data complexity may lead to diminished accuracy even though the dataset has smaller object count.

**Learning with BDNN transfers previously learned knowledge**

As the development of a *f*-BDNN always starts from scratch when an input dataset is given, different *f*-BDNNs are independent on each other. In Fig 3A to C, three *f*-BDNNs were developed upon dataset $X(O_{10})$, $X(O_{15})$ and $X(O_{20})$ (Fig. S2), respectively, where $O_M$ represents a class of M different objects. Here, $X(O_{10}) \subset X(O_{15}) \subset X(O_{20})$. One observed that the *f*-BDNN($O_{10}$) reached the highest test accuracy, 96%, it only took 0.22 hours (≈13 min) with 220 hidden neurons in the network (Fig. 3A). It took 1.5 hours with 300 hidden neurons for *f*-BDNN($O_{15}$) (Fig. 3B), and 1.8 hours with 410 hidden neurons for *f*-BDNN($O_{20}$) (Fig. 3C) to reach 90% of test accuracy.

To assess the capability of BDNN in knowledge transfer, as shown in Fig. 3D, an *e*-BDNN($O_{10}$) was constructed upon the input dataset $X(O_{10})$ by starting from an *f*-BDNN($O_5$) constructed upon a dataset $X(O_5)$ where $X(O_5) \subset X(O_{10})$. In the other word, the *f*-BDNN($O_5$) works as the seed of *e*-BDNN($O_{10}$). Initially, the *e*-BDNN($O_{10}$) has the same number of hidden neurons as the *f*-BDNN($O_5$), i.e.,63 hidden neurons. With the 63 hidden neurons in the network, the *e*-BDNN($O_{10}$) underwent one loop of configuration of synapses upon the input dataset $X(O_{10})$, which changed the number of output neurons from 5 to 10. Surprisingly, this network has been already able to classify the 10 objects in $O_{10}$ with the accuracy as high as 89%. Subsequently, the *e*-BDNN starts to grow, recruiting incrementally new hidden neurons upon $X(O_{10})$. The training and testing accuracies were continuously increased and reached the 98% within only 0.17 hour (≈10 min) when 150 more hidden neurons were recruited into the network (totaling 213).

To further show the capability of knowledge transfer, the *e*-BDNN($O_{10}$) with 98% accuracy was used as the seed to learn the larger dataset $X(O_{15})$ (Fig. 3E). Similarly, the initial *e*-BDNN($O_{15}$) can classify rather well the $O_{15}$ with 91% test accuracy. The growth of the *e*-BDNN($O_{15}$) increased test accuracy to 93% within 0.17 hour when 50 more hidden neurons were recruited into the network (totaling 267). Subsequently, the *e*-BDNN($O_{15}$) with 93% accuracy became a seed to learn the dataset $X(O_{20})$ (Fig. 3F). Fed with the dataset $X(O_{20})$, the initial *e*-BDNN($O_{15}$) has the accuracy of 88%, and was increased to 90% by incorporating 110 more hidden neurons into the network.

The results (Fig. 3D to F) show clearly that our BDNN-based learning transferred previously learned knowledge, thus overcoming the catastrophic forgetting that the BP-based learning procedure suffers from. The high-level knowledge transfer substantially facilitates the learning process upon the larger size of dataset containing incremental new objects with the learning speed 10 times faster than the fresh learning of the same dataset (Fig. 3A to C).

**Learning with BDNN is beyond the backpropagation procedure**

With the similarly high classification accuracies upon the same input datasets, learning with BDNN is substantially beyond the BP-based learning procedure in terms of the automation, generalization and learning speed.

Firstly, it is known the BP-based learning procedure requires hyperparameter modulation which normally involves manual operations by experienced engineers. Examples of hyperparameters to be determined include the number of hidden layers and the number of hidden neurons in each layer. Programmed search of hyperparameters, such as the Optuna framework (*36*), is possible to facilitate the hyperparameter modulation, but at a cost of large amount of computation. As a contrast, our learning approach with BDNN rules out the need of hyperparameter modulation during the learning process, and additionally, the development of BDN is automatically adapted to variations in input different datasets (Fig. 2).

Secondly, our learning BDN is superior in the generalization to new data as a result of the inherent capability of knowledge transfer. The knowledge transfer prevents BDNNs from the catastrophic forgetting encountered by the BP-based learning method (Fig. 4A). To compare the generalization of the BDNN with BP-SNN and BP-CSNN (convolutional spiking neural network), they had completed training upon $X(O_5)$. Subsequently, they were trained for only one-loop towards an enlarged dataset incorporated five more new objects $X(O_{10})$, i.e., $X(O_{10}) \supset X(O_5)$. The knowledge transfer of the BDNN immediately results in the high accuracies of 89% in classifying $O_{10}$. It is analog to that a child, after grasping each toy merely for one time, can more quickly recognize all toys with new toys added to those the child has played previously. As a comparison, the BP-based learning procedures show the typical catastrophic forgetting as the accuracies were merely 59% and 48% for the BP-CSNN and BP-SNN, respectively, upon the enlarged dataset $X(O_{10})$ (Fig. 4A and S4).

Thirdly, to evidence the substantially faster learning of the BDNNs over the BP-based learning, comparison in running different algorithms on the same hardware upon the same dataset $O_{20}$ was conducted. The algorithms include the *f*-BDNN and *e*-BDNN, and the BP-based learning procedure with SNN and CSNN. For the latter, surrogate function was used to enable the gradient calculations. Assuming an engineer is highly experienced, one guess of the number of hidden layers and the number of hidden neurons in a hidden layer works. In this ideal case, the total length of learning time equals to the length of training time. In real cases, however, multiple trials in the hyperparameter modulation are common. As a consequence, the total length of learning time is the sum of durations of multiple training trails.

One can clearly observe in Fig. 4B that the lengths of the learning time for the BDNNs are substantially shorter than those of BP-based learning procedures in any cases when similar classification accuracies around 90% were reached. In the detail (SM), as displayed in Table S2, it took 1.9, and 0.7 hour for *f*-BDNN and *e*-BDNN to reach this level of accuracy. To reach the highest accuracies they can, the lengths of the training time were increased to 9 and 6 hours, respectively. As a comparison, the learning speed is more than 15 times lower as it took 29 and 13 hours for training BP-SNN and BP-CSNN when the right number of neurons was set, which was the ideal case. In the real cases when the modulation of the number of hidden layers and the number of hidden neurons in each layer was involved, the length of learning time was multiplied as the consequence (Fig. 4B, Table S2).

**References and Notes**
1. K. Yamazaki, V.-K. Vo-Ho, D. Bulsara, N. Le, Spiking Neural Networks and Their Applications: A Review. *Brain Sci.* **12** (2022).

2. D. Li, X. Chen, M. Becchi, Z. Zong, "Evaluating the Energy Efficiency of Deep Convolutional Neural Networks on CPUs and GPUs" in *2016 IEEE International Conferences on Big Data and Cloud Computing (BDCloud), Social Computing and Networking (SocialCom), Sustainable Computing and Communications (SustainCom) (BDCloud-SocialCom-SustainCom)* (2016; https://ieeexplore.ieee.org/document/7723730), pp. 477–484.

3. N. Rathi, K. Roy, DIET-SNN: A Low-Latency Spiking Neural Network With Direct Input Encoding and Leakage and Threshold Optimization. *IEEE Trans. Neural Netw. Learn. Syst.* **34**, 3174–3182 (2023).

4. K. Malcolm, J. Casco-Rodriguez, A Comprehensive Review of Spiking Neural Networks: Interpretation, Optimization, Efficiency, and Best Practices. arXiv arXiv:2303.10780 [Preprint] (2023). https://doi.org/10.48550/arXiv.2303.10780.

5. B. Wu, X. Dai, P. Zhang, Y. Wang, F. Sun, Y. Wu, Y. Tian, P. Vajda, Y. Jia, K. Keutzer, "FBNet: Hardware-Aware Efficient ConvNet Design via Differentiable Neural Architecture Search" (2019; https://openaccess.thecvf.com/content_CVPR_2019/html/Wu_FBNet_Hardware-Aware_Efficient_ConvNet_Design_via_Differentiable_Neural_Architecture_Search_CVPR_2019_paper.html), pp. 10734–10742.

6. B. Han, K. Roy, "Deep Spiking Neural Network: Energy Efficiency Through Time Based Coding" in *Computer Vision – ECCV 2020*, A. Vedaldi, H. Bischof, T. Brox, J.-M. Frahm, Eds. (Springer International Publishing, Cham, 2020), pp. 388–404.

7. S. Li, Z. Zhang, R. Mao, J. Xiao, L. Chang, J. Zhou, A Fast and Energy-Efficient SNN Processor With Adaptive Clock/Event-Driven Computation Scheme and Online Learning. *IEEE Trans. Circuits Syst. Regul. Pap.* **68**, 1543–1552 (2021).

8. Y. Hu, Q. Zheng, X. Jiang, G. Pan, Fast-SNN: Fast Spiking Neural Network by Converting Quantized ANN. *IEEE Trans. Pattern Anal. Mach. Intell.* **45**, 14546–14562 (2023).

9. A. Zhang, X. Li, Y. Gao, Y. Niu, Event-Driven Intrinsic Plasticity for Spiking Convolutional Neural Networks. *IEEE Trans. Neural Netw. Learn. Syst.* **33**, 1986–1995 (2022).

10. Y. Wang, M. Zhang, Y. Chen, H. Qu, "Signed Neuron with Memory: Towards Simple, Accurate and High-Efficient ANN-SNN Conversion" in *Proceedings of the Thirty-First International Joint Conference on Artificial Intelligence* (International Joint Conferences on Artificial Intelligence Organization, Vienna, Austria, 2022; https://www.ijcai.org/proceedings/2022/347), pp. 2501–2508.

11. An exact mapping from ReLU networks to spiking neural networks - ScienceDirect. https://www.sciencedirect.com/science/article/pii/S0893608023005051.

12. A Tandem Learning Rule for Effective Training and Rapid Inference of Deep Spiking Neural Networks | IEEE Journals & Magazine | IEEE Xplore. https://ieeexplore.ieee.org/abstract/document/9492305?casa_token=7D4YQ4dndNYAAAAA:a_0fSVsWct6WoVD2lnr2Tf1JKfTqlvQXgJd-V4q0Fr7oxgm-NR_Jh8IgMYRTgXzY2coscFME.

13. Y. Wang, H. Liu, M. Zhang, X. Luo, H. Qu, A universal ANN-to-SNN framework for achieving high accuracy and low latency deep Spiking Neural Networks. *Neural Netw.*, 106244 (2024).

14. E. O. Neftci, H. Mostafa, F. Zenke, Surrogate Gradient Learning in Spiking Neural Networks. arXiv arXiv:1901.09948 [Preprint] (2019). https://doi.org/10.48550/arXiv.1901.09948.

15. J. Shen, Q. Xu, J. K. Liu, Y. Wang, G. Pan, H. Tang, ESL-SNNs: An Evolutionary Structure Learning Strategy for Spiking Neural Networks. *Proc. AAAI Conf. Artif. Intell.* **37**, 86–93 (2023).

16. S3NN: Time step reduction of spiking surrogate gradients for training energy efficient single-step spiking neural networks - ScienceDirect. https://www.sciencedirect.com/science/article/pii/S0893608022005007.

17. B. M. Lake, M. Baroni, Human-like systematic generalization through a meta-learning neural network. *Nature* **623**, 115–121 (2023).

18. J. A. Fodor, Z. W. Pylyshyn, "Connectionism and Cognitive Architecture" in *Readings in Philosophy and Cognitive Science*, A. I. Goldman, Ed. (The MIT Press, 1993; https://direct.mit.edu/books/book/3952/chapter/165126/Connectionism-and-Cognitive-Architecture).

19. A. J. Nam, J. L. McClelland, Systematic human learning and generalization from a brief tutorial with explanatory feedback. arXiv arXiv:2107.06994 [Preprint] (2023). http://arxiv.org/abs/2107.06994.

20. T. Wang, P. Zheng, S. Li, L. Wang, Multimodal Human–Robot Interaction for Human-Centric Smart Manufacturing: A Survey. *Adv. Intell. Syst.* **n/a**, 2300359.

21. F. Liu, S. Deswal, A. Christou, Y. Sandamirskaya, M. Kaboli, R. Dahiya, Neuro-inspired electronic skin for robots. *Sci. Robot.* **7**, eabl7344 (2022).

22. A. J. Bremner, C. Spence, "Chapter Seven - The Development of Tactile Perception" in *Advances in Child Development and Behavior*, J. B. Benson, Ed. (JAI, 2017; https://www.sciencedirect.com/science/article/pii/S0065240716300477)vol. 52, pp. 227–268.

23. S. Pyo, J. Lee, K. Bae, S. Sim, J. Kim, Recent Progress in Flexible Tactile Sensors for Human-Interactive Systems: From Sensors to Advanced Applications. *Adv. Mater.* **33**, 2005902 (2021).

24. B. Deng, Y. Jiao, T. Gao, G. Yi, J. Wang, J. Wang, "Object Tactile Classification Based on Convolutional Spiking Neural Network with Attention Mechanism" in *2023 42nd Chinese Control Conference (CCC)* (2023; https://ieeexplore.ieee.org/abstract/document/10240699), pp. 8258–8263.

25. J. Lee, J. Y. Kwak, K. Keum, K. Sik Kim, I. Kim, M.-J. Lee, Y.-H. Kim, S. K. Park, Recent Advances in Smart Tactile Sensory Systems with Brain-Inspired Neural Networks. *Adv. Intell. Syst.* **n/a**, 2300631.

26. H. Niu, H. Li, S. Gao, Y. Li, X. Wei, Y. Chen, W. Yue, W. Zhou, G. Shen, Perception-to-Cognition Tactile Sensing Based on Artificial-Intelligence-Motivated Human Full-Skin Bionic Electronic Skin. *Adv. Mater.* **34**, 2202622 (2022).

27. L. Chen, S. Karilanova, S. Chaki, C. Wen, L. Wang, B. Winblad, S.-L. Zhang, A. Özçelikkale, Z.-B. Zhang, Spike timing–based coding in neuromimetic tactile system enables dynamic object classification. *Science* **384**, 660–665 (2024).

28. S. Micera, Toward more naturalistic tactile sensors. *Science* **384**, 624–625 (2024).

29. E. O. Neftci, H. Mostafa, F. Zenke, Surrogate Gradient Learning in Spiking Neural Networks: Bringing the Power of Gradient-Based Optimization to Spiking Neural Networks. *IEEE Signal Process. Mag.* **36**, 51–63 (2019).

30. H. Wu, Y. Zhang, W. Weng, Y. Zhang, Z. Xiong, Z.-J. Zha, X. Sun, F. Wu, Training Spiking Neural Networks with Accumulated Spiking Flow. *Proc. AAAI Conf. Artif. Intell.* **35**, 10320–10328 (2021).

31. D. Purves, G. J. Augustine, D. Fitzpatrick, W. Hall, A.-S. LaMantia, L. White, *Neurosciences* (De Boeck Superieur, 2019).

32. B. P. Graham, A. van Ooyen, Mathematical modelling and numerical simulation of the morphological development of neurons. *BMC Neurosci.* **7**, S9 (2006).

33. Neuronal Dynamics - a neuroscience textbook by Wulfram Gerstner, Werner M. Kistler, Richard Naud and Liam Paninski. https://neuronaldynamics.epfl.ch/.

34. The Postnatal Development of the Human Cerebral Cortex. Vol. 1. The Cortex of the Newborn. *J. Anat.* **73**, 674 (1939).

35. I. Goodfellow, Y. Bengio, A. Courville, *Deep Learning* (MIT Press, 2016).

36. T. Akiba, S. Sano, T. Yanase, T. Ohta, M. Koyama, "Optuna: A Next-generation Hyperparameter Optimization Framework" in *Proceedings of the 25th ACM SIGKDD International Conference on Knowledge Discovery & Data Mining* (Association for Computing Machinery, New York, NY, USA, 2019; https://dl.acm.org/doi/10.1145/3292500.3330701)*KDD '19*, pp. 2623–2631.

**Acknowledgments:** This work was partially supported by the Swedish Foundation for Strategic Research (FUS21- 0067) and the Swedish Research Council (2019-05484).

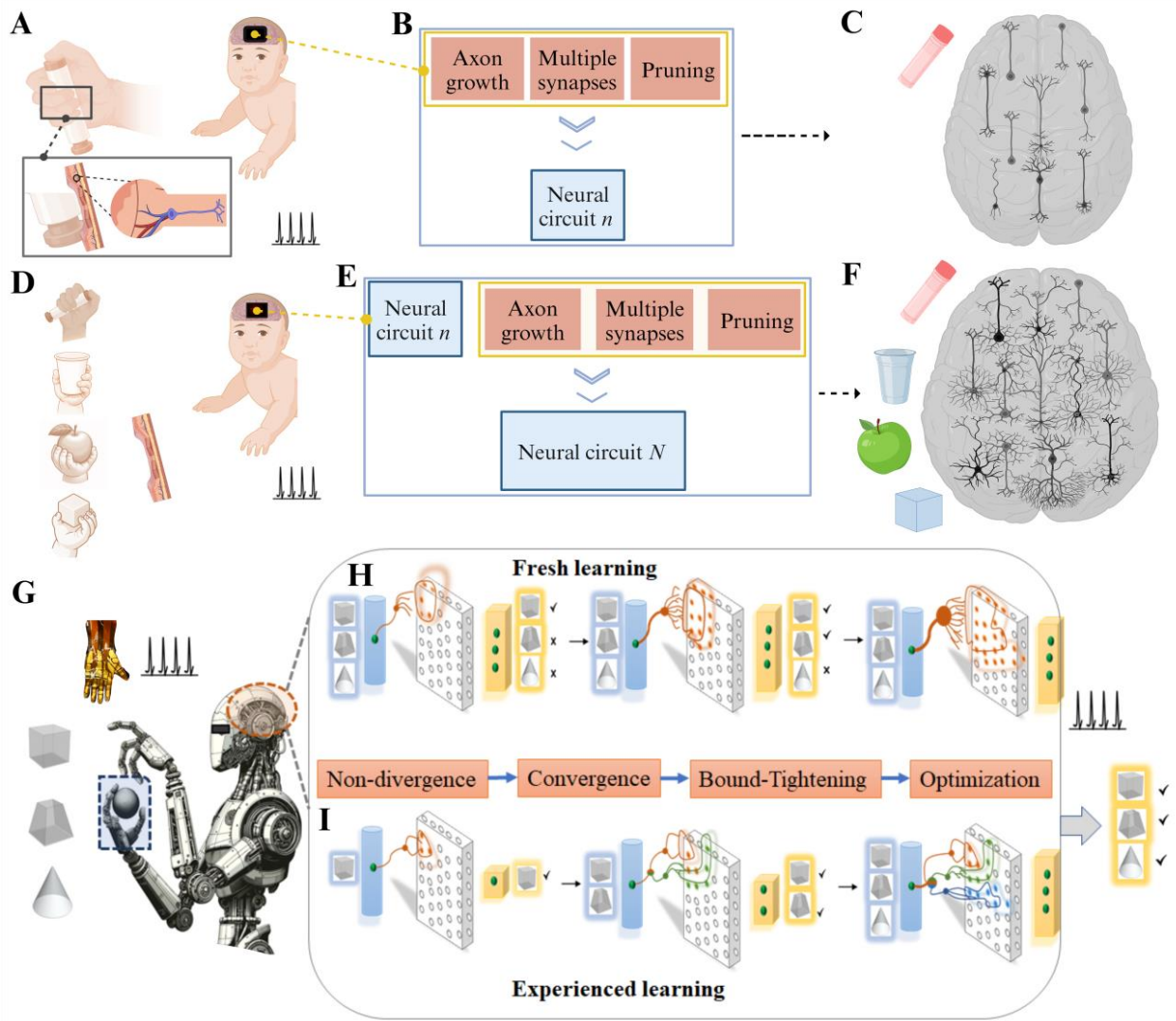

**Fig.1. Schematic illustration of (A-F) developmental neural circuits for tactile perception and (G-I) the inspired learning model with brain-mimetic developmental spiking neural networks (BDNN) used in a neuromorphic tactile system**. (A), a child learns, for the first time, a cylindric tube by grasp that activates mechanoreceptors in the hand generating spike trains. (B) the spike trains are delivered to the brain to recruit neurons by stimulating neuronal growth, multiple synapse formation and pruning, (c) ultimately forming neural circuit *n* that represent the concept of the cylindric tube. (D) The previously learned knowledge represented in neural circuit *n* facilitates the child to learn new objects (cubic, cup and apple) (E) where the neural circuit *n* is used and incorporated into the enlarged neural network *N* (F) that represent all the objects. (G) The neuromorphic tactile system employed spike trains generated by using a neuromorphic e-skin in grasping objects to drive learning processes with BDNNs. (H) In a fresh learning, a BDNN is developed from scratch upon the dataset of cubic, trapezoid, and cone by incrementally recruiting hidden neurons until accurate classification of the objects is reached. (I) In an experienced learning, a BDNN learns the cubic firstly. The BDNN for cubic is used and incorporated into the BDNN for the cubic and the trapezoid. The latter BDNN is similarly used and incorporated into the BDNN for the cubic, the trapezoid and the cone. Illustration was created with BioRender.com.

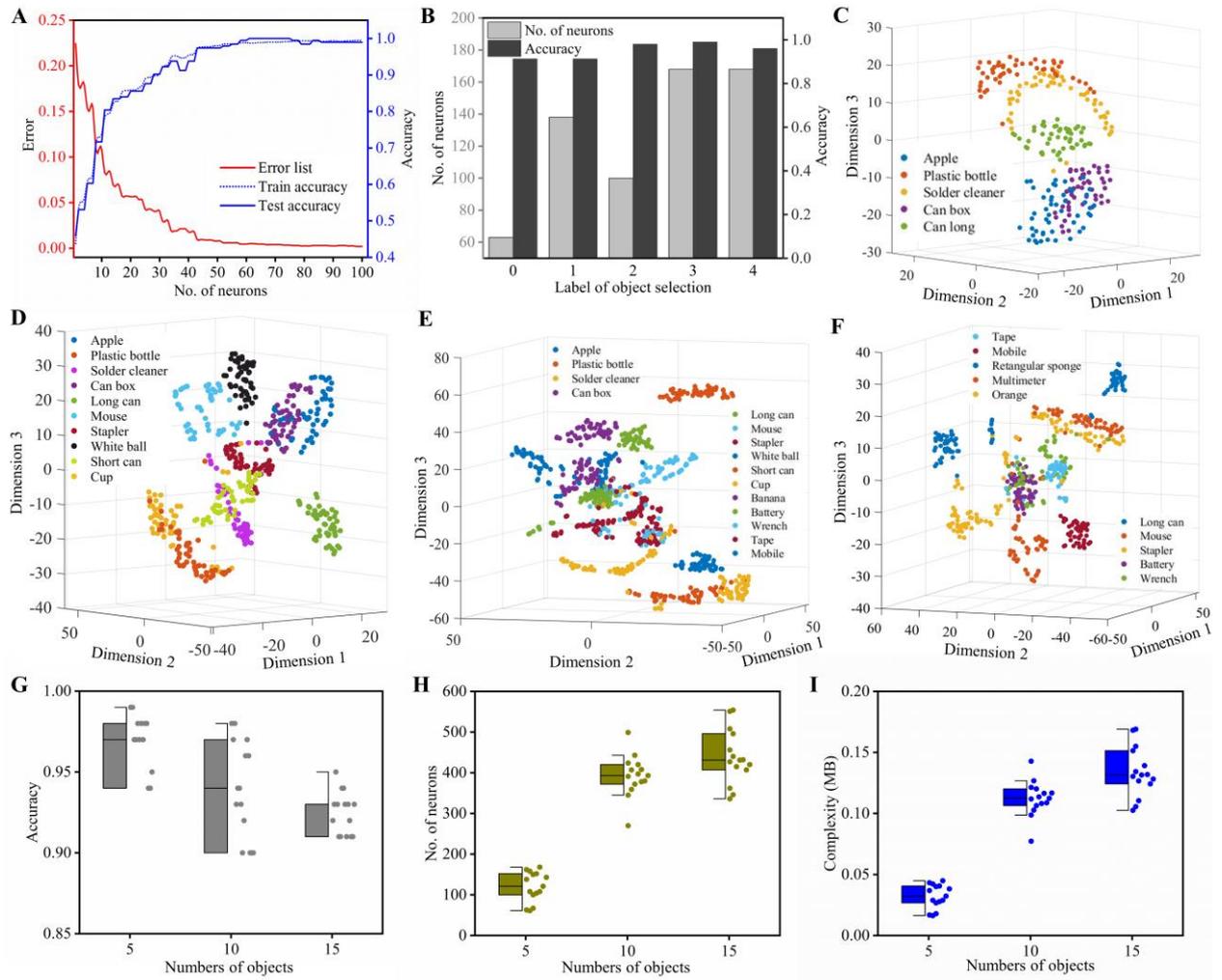

**Fig. 2. Fresh learning with BDNNs by grasp**. (A) The learning process with *f*-BDNN upon a 5-class dataset (Class 5-0) where the error and accuracy changes with the number of hidden neurons. (B) Five *f*-BDNNs constructed upon the five different 5-class datasets (label 0 to 4), separately, have different number of hidden neurons to reach the highest accuracies. Feature distribution of (C) the 5-class (Class 5-0), (D) a 10-class, and (E) a 15-class dataset visualized by t-SNE. (F) Feature distribution of another 10-class dataset containing different types of objects from (D). (G) Accuracies, (H) required numbers of hidden neurons and (I) space complexities of the *f*-BDNNs that learned all sets of objects, separately, where the sets are grouped into 5, 10 and 15 classes.

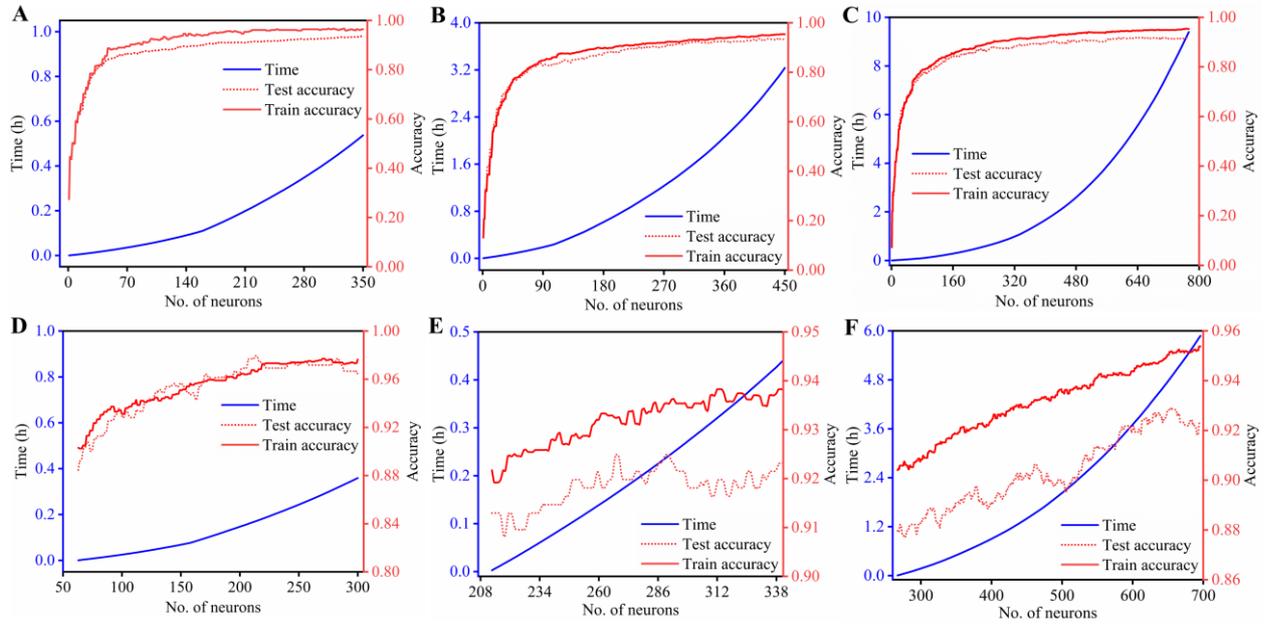

**Fig. 3. Comparison between (A-C) fresh learning and (D-F) experienced learning**. Spent time and accuracies as a function of the number of recruited hidden neurons of (A) a *f*-BDNN($O_{10}$) upon a 10-class dataset $X(O_{10})$, (B) a *f*-BDNN($O_{15}$) upon a 15-class dataset $X(O_{15})$, and (C) a *f*-BDNN($O_{20}$) upon the 20-class dataset $X(O_{20})$. Spent time and accuracies as a function of the total number of recruited hidden neurons of (D) an *e*-BDNN($O_{10}$) upon the dataset $X(O_{10})$ by commencing from a *f*-BDNN($O_5$) trained using $X(O_5)$ where $X(O_5) \subset X(O_{10})$ with the highest accuracy of 97%, (E) an *e*-BDNN($O_{15}$) upon the 15-class dataset $X(O_{15})$, by commencing from the *e*-BDNN($O_{10}$) with the highest accuracy 98% in (D), and (F) an *e*-BDNN($O_{20}$) by commencing from the *e*-BDNN($O_{15}$) with the highest accuracy 93% in (E).

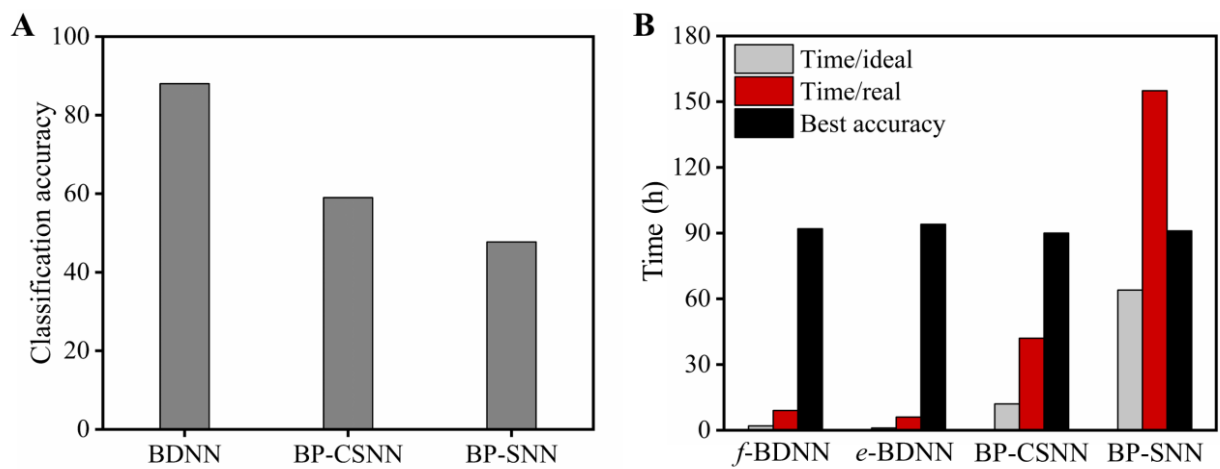

**Fig. 4**. **Comparison in the (A) generalization and (B) learning speed between BDNN-based and the BP-based learning approaches**. All algorithms were run on the same hardware, AMD Ryzen 7 5800hz, for learning gasp of the 20 objects $O_{20}$. Time/ideal refers to the ideal application environment of the BP-based learning procedure where only one step of manual modulation of network structure was conducted to get the best learning component with the shortest time length. In this case, a highly experienced engineer is required. Time/real refers to real application situations of the BP-based procedure where multiple-step manual structure modulation was required. Here, three steps for BP-SNN, and four steps for BP-CSNN (see Table S2) were exemplified.

**Materials and Methods**

I. The input datasets generated by grasp objects

The datasets were generated by using our in-lab neuromorphic electronic (e-) skin comprising 64 tactile sensors, connected with a neuromorphic electronic circuit which were fabricated based on printed circuit board (PCB) technology. The detail of data collection was described in our previous report (27). In brief, by using the e-skin, grasp of an individual object were repeated 50 times, generating 50 data samples for each object. Each data sample comprises 64 data streams as there are 64 sensors in the e-skin. The data streams are in the form of spike trains. A spike has amplitude of 2 V and width 1.5 ms. A 20-object pool consists of 20 different objects in total, including apple, plastic bottle, solder cleaner, can box, long can, mouse, stapler, white ball, short can, cup, banana, battery, tape, mobile, rectangular sponge, multimeter, spray, ball tennis and orange. (Fig S2). There were four different participants who involved in grasping the objects. It is noted that there relatively large variations in data samples from grasp to grasp, and from participant to participant. For one object, there are 200 data samples. In Each data sample, there are 1600 data points.

To show the adaptability of the learning with BDNN, we randomly selected five different objects from a pool of 20 objects for grasp, creating a 5-class dataset which is denoted as Class 5-0. By repeating the selection for four more times, four more 5-class datasets were generated, Thus, we have five 5-class datasets, denoted as Class 5-$i$ ($i$=0, 1, $\cdots$, 4) respectively.

Subsequently, to Class 5-0, data samples of five objects which were randomly selected from the rest of the pool were added, forming a 10-class dataset, denoting it as Class 10-0. The same way was taken to form Class 10-$i$ from Class 5-$i$, where $i$ spans from 1 to 4. Similarly, 15-class datasets were formed from the 10-class datasets, i.e., Class 15-$i$ from Class 15-$i$ ($i$=0, 1, $\cdots$4). For instance, data samples of five objects which were randomly selected from the rest of the pool were added to Class 10-0, forming Class 15-0 (Fig.S2)

As a result of the selection procedure, we have 16 datasets in total, comprising five 5-class datasets, five 10-class datasets, five 15-class datasets, and one 20-class dataset which is the pool.

II. Learning experiments and assessment

For each of the 16 datasets, fresh learning with BDNN was repeated three times. Therefore, we have created 15 $f$-BDNNs for the 5-class datasets, the 10-class datasets, and the 15-class datasets, respectively, and three $f$-BDNNs for the 20 objects. In total, we have hence created 45 $f$-BDNNs. For the purpose of comparison, the learning approach with BDNN is benchmarked in the learning speed with the BP-based learning method with SNN and CSNN reported lately (29,30). All the algorithms for the comparison were trained using the same dataset of the 20 objects.

To assess the generalization of BDNN, an $e$-BDNN which has conducted only one-loop learning was examined. Upon an enlarged dataset, one-loop learning leads to the e-BDNN which has the same number of hidden neurons as its prior network. Simultaneously, the one-loop learning has increased the number of output neurons in the *e*-BDNN required by the enlarged dataset. For instance, the BDNN which has been already constructed upon Class 5-0, when fed with Class 10-0, underwent one-loop of the learning process. The number of the hidden neurons does not change while the number of output neurons is increased to from 5 to 10 in the one-loop learning. The same procedure was applied when the BDNN upon Class 10-0 was used as the seed to construct the BDNN upon Class 15-0.

All algorithms developed in this work were implemented on a laptop equipped with an AMD Ryzen 7 5800X processor. For all the BP-based learning algorithms, the models at the last epoch (i.e., the 100$^{th}$ ep.) were saved.

III. The learning theory

*1. The bio-inspiration*

Our BDNN emulates the development of neural circuits in early brain with several key features (29). Firstly, information about experiences is represented, processed, and transmitted mainly in action potential, i.e., spikes. Secondly, driven by neuronal activities associated with experiences, axons seek and find pathways to create valuable synapses with varying synaptic strength. Thirdly, pruning neurons with un-valuable synapses occurs simultaneously. The combination of the pathfinding and pruning processes, which happen multiple times, leads to diverse network architectures representing knowledge of objects from learning processes. Lastly, an existing neural circuit representing previously learned knowledge is effectively used to learn new objects (17).

As referred to Fig S1, mathematical convergence is ensured in the construction of BDNN. Here, the LIF neuron model was applied. In the pathfinding unit, we randomly choose different weight combinations to define a pathfinding set. The weights are examined individually by using an index $\boldsymbol{\xi_n}$. Those synapses with their weights satisfying convergence are selected as valuable pathfinding solutions. From the set of solutions, in the pruning unit, the weights which satisfy a bond-tightening condition are further selected. The continuity in the growth of hidden layer neurons is determined by the performance requirement of a learning task, e.g., classification accuracy. The weights of the output layer are obtained by the least square solution.

In learning new and more objects, the synaptic weights of the input and hidden layer in the experience unit remains unchanged. In this experienced learning, the BDNN decides whether to continue the growth of neurons of the hidden layer by comparing the output of the existing basic network component, which is smaller, and the expected learning result. The existing basic network component facilitates to select and calculate the weights of the BDNN.

*2. The input-output relation*

Given a dataset of spike trains of $a$ objects $X(O_a) \in N_a \times T \times d$, $X(O_a)(i) = [x_a(1), x_a(2), \ldots, x_a(T)], i = 1,2 \ldots N_a$, where $O_a, N_a, T, d$ represent the set of the $a$ objects, the number of data samples, the length of data streams, and the number of tactile sensors (64 in our case). For an enlarged set of $b$ objects which includes the set of the $a$ objects, $b = a + a', a' > 0$ where $a'$ is the number of new objects from the rest of the 20-object pool. The dataset of $b$ objects is denoted as $X(O_b) \in N_b \times T \times d$, and satisfies

$$X(O_b) = X(O_a) \cup X(O_{a'})$$

The target function of BDNN is

$$f : R^d \rightarrow R^m$$

where $m$ is the number of objects to be classified in a task. In BDNN, we suppose that there is one learning component. This learning component consists of the set of the synapses between the input and the hidden layer.

Supposed that a BDNN has learned $X(O_a)$ from scratch (i.e., *f*-BDNN) where there are $n-1$ hidden neurons in the network, the learning component is characterized by the weights $W_{a_{n-1}}$. If the accuracy of the *f*-BDNN has not reached a desired level, it continuously grows by adding $n-1$ hidden neurons in sequence to the network where the integer $n \geq 1$. The output of the *f*-BDNN is

$$f_{n-1}(x) = \sum_{j=1}^{n-1} \beta_{n-1} S_j \qquad (1)$$

where $x \in X(O_a)$, $S_j$ is the spike train from the $j$th hidden neuron, $\beta_{n-1}$ is weights between the $n-1$ incremented hidden neurons and the output neurons. The residual fitting error is denoted as $e_{n-1}$

$$e_{n-1} = f - f_{n-1}(x) = f - \sum_{j=1}^{n-1} \beta_{n-1} S_j \qquad (2)$$

where $x \in X(O_a)$.

Supposed the BDNN has completed learning $O_a$ with $n$ hidden neurons in the network, an enlarged dataset $X(O_b)$ appears. This BDNN is used to learn $X(O_b)$ and hence continues to grow which is the experienced learning (*e*-BDNN). Similarly, supposed that the resulting *e*-BDNN has $n-1$ new incremental hidden neurons, the output of the *e*-BDNN is

$$f_{n-1}(x) = \sum_{j=1}^{n+n-1} \beta_{n-1} S_j \qquad (3)$$

where $x \in X(O_b)$. The residual fitting error $e_{n-1}$ is

$$e_{n-1} = f - f_{n-1}(x) = f - \sum_{j=1}^{n+n-1} \beta_{n-1} S_j \qquad (4)$$

It is noted that an *e*-BDNN with $n=1$, $e_0 = f - \sum_{j=1}^{n} \beta_1 S_j$. It deserves to stress that this *e*-BDNN only undergoes one-loop learning when being fed with the enlarged dataset $X(O_b)$.

*3. The node component*

The development of the $n$th new hidden neuron in the node component is implemented by the path-finding unit and pruning unit, where $n$ is a positive integer. If the learning accuracy of the BDNN has not reached the desired level, the pathfinding unit works in a way that the axon of a pre-synaptic neuron (i.e., input neuron) attempts to establish connections with the dendrites of $P$ neurons in the hidden layer. Subsequently, in the pruning unit, the post-synaptic hidden neurons with optimal weights ($w_n^*$ and $v_n^*$) are selected from these $P$ pathfinding attempts. Those selected hidden neurons are hence suitable for translating the information from the input neurons.

For each pathfinding operation, distinct synaptic strengths are characterized by weights as $w_n$ and $v_n$, where $w_n$ denotes the weights between the input neurons with the hidden neuron $n$, and $v_n$ is the weight of output feedback of hidden neuron $n$. They can be originally defined by using random operations. In the $P$ times of pathfinding operations, the resulting $w_n$ and $v_n$ can be denoted as $[-\lambda, \lambda]^{d \times P}$ and $diag[-\lambda, \lambda]^P$, respectively. In LIF neuronal dynamics, the input current to a neuron is typically the integration of synaptic currents triggered by the arrival of presynaptic spikes. We assume that the synaptic currents can be summed linearly. Assuming the time step of spike trains is $\Delta t$, $S_n$, $S_n = \{S_{n_1}; S_{n_2}; \ldots, S_{n_p}; \ldots; S_{n_P}\}$ represents the output of the $n$ th newly incremented hidden neuron, where $S_{n_p}\left(X(O_b), w_{n_p}, v_{n_p}\right) =$

$\left[s_{n_p}(1), \ldots, s_{n_p}(t), \ldots, s_{n_p}(T)\right], p = 1, 2, \ldots, P$. $S_{n_p}$ is obtained by using eq. 5 and 6, where $\tau_{syn}$ and $\tau_{mem}$ are synaptic and membrane time constants, and $\Theta$ is the Heaviside step function.

$$\begin{cases} i_{n_p}(t) = \exp(-(\Delta_t/\tau_{syn})) \cdot i_{n_p}(t-1) + w_{n_p} \cdot x(t) + v_{n_p} \cdot s_{n_p}(t-1) \\ u_{n_p}(t) = \exp(-(\Delta_t/\tau_{mem})) \cdot u_{n_p}(t-1) + i_{n_p}(t-1) - s_{n_p}(t-1) \end{cases}$$
(5)

$$s_{n_p}(t) = \Theta(u_{n_p}(t) - \vartheta)$$
(6)

where $t = 1, 2, \ldots, T$. As $\|e_n\|^2 = \|f - f_{n-1} - \beta_n S_n\|^2$, the square of the error $\|e_n\|^2$ is calculated as $\langle e_{n-1} - \beta_n S_n, e_{n-1} - \beta_n S_n \rangle$. One only considers the influence of the weights between the input and the hidden layer on the learning in the node component. Under the condition, our BDNN can quantify the non-divergent of the learning such that $\|e_n\|^2 \leq \|e_{n-1}\|^2$. However, the BDNN cannot quantify that each pathfinding is useful for the learning task.

In the pruning unit, one index $\xi_n$ is defined as $\xi_n = \xi_n(e_{n-1}, S_n, \sigma)$, where $0 < \sigma < 1$. It is used to retain the right paths during the pathfinding and also to ensure the convergence such that $\|e_n\|^2 < \|e_{n-1}\|^2$ by defining the range of the index. The associated $w_n$ and $v_n$ in the right paths are saved in $W_n$.

Subsequently, the pruning unit selects the optimal $w_n^*$ and $v_n^*$ to make the upper-limit value $\varepsilon$ of $\|e_n\|^2$ be minimum by the operation of $\max(\xi_n)$, where $\varepsilon > 0$. The calculation of $e_{n-1}$ in $\xi_n$ of $e$-BDNN is conducted by using $f - \sum_{j=1}^{\mathfrak{n}+n-1} \beta_{n-1} S_j$, when $1 \leq j \leq \mathfrak{n}$, $S_j = S_j(X(O_b), W_{a_n})$, and $S_j = S_j(X(O_b), W_{n-1}^*)$ when $1 + \mathfrak{n} \leq j \leq \mathfrak{n} + n - 1$. Assuming each convergent pathfinding is non-optimal with a probability $\eth$, where $\forall \eth \in (0, 1)$, the probability of the optimal pathfinding $w_n^*$ and $v_n^*$ from $W_n$ approaches $1 - \eth^P$ which is close to 1.

    *4. Wrench component*

The wrench component decodes the tactile information in the hidden layer and further transform it into output which corresponds to labels of objects. In terms of operation, the wrench component aims to obtain suitable weights between the hidden and the output layer by experienced unit and docking unit.

Experienced unit concerns the optimal weights $W_n^*$, $W_n^* = [w_1^*, w_2^*, \ldots, w_n^* | v_1^*, v_2^*, \ldots, v_n^*]$ of the input neurons to the recruited $n$ hidden neurons. During the development, the weights $\mathbb{W}_n = [W_{a_n}; W_n^*] \in R^{d \times (\mathfrak{n}+n)}$ of the input neurons to the $\mathfrak{n} + n$ hidden neurons. Fitting index is illustrated with $\|e_n\|^2$, where $S_n^*$, $S_n^* = S(X, \mathbb{W}_n)$ is the output of all recruited hidden neurons. The function of the docking unit is to update $\beta$ by using $\min(\|e_n\|^2)$ which is calculated by the derivative solving method.

    <u>On benchmarking with the BP-based learning procedure</u>

    In addition to the advantage of our learning approach that the BDNN does not require network modulation, our BDNN is superior over the BP-based learning procedure in terms of the learning speed and the generalization as shown in Fig 3, 4, and Fig S4. For the comparison of the two learning approaches, all the algorithms used the LIF spiking neuron model and used

the same input dataset of the 20-class objects. All the codes were run on the same hardware (Method). Here, the number of hidden neurons is the relevant hyper-parameter of the network. We run the BP-based procedures for 100 epochs and looked for the epoch which produces the highest accuracy. As shown in Table. S2, for the three-layer SNN with 50 hidden neurons (BP-SNN-1), the accuracy reached 90%. It took 29.1 hrs to run the training process for 100 epochs. That is, one has 0.29 hr/epoch. One experienced engineer may predict that 90% is the highest accuracy the training process can reach. In this case, the program can be set so that the training is stopped once the testing accuracy of 90% is reached. In this case (67 epochs), the training time can be reduced to 19.5 hrs.

In order to achieve the highest accuracy from the BP-based learning, one has to repeat the training process by tuning, often manually, the hyper-parameters. The learning time to achieve such a best network is the total time of all epochs when the hyperparameters are tuned. If the engineer tried firstly one hidden layer, followed by adding a second hidden layer (BP-SNN-2), the total training time can be 59 hrs. Alternatively, if one tried one hidden layer with 50 neurons, followed by adding 50 more hidden neurons, the resultant BP-SNN-3 has an accuracy of 91% but costed 155 hrs for the training. One may continue to repeat the training a few more times by increasing the number of hidden neurons further for possible higher accuracies. In practice, one likely tunes the number of hidden neurons by starting from e.g., 10, added by e.g., 10 or 20 each time. In any case, the learning time is multiplied by the substantially increased number of epochs. The similar results were obtained when benchmarking with the BP-based learning with convolutional spiking neural networks (CSNN).

In contrast, our BDNNN-based learning, as the fundamentally different learning mechanism form the BP-based learning, leads to the substantially reduced leaning time to reach the same accuracy. It takes only 1.94 and 9.4 hrs for the *f*-BDNNN fed with the 20-class object dataset to reach 90 and 92% accuracies, respectively. The learning time for the *e*-BDNN is significantly shorter, i.e., which is 0.73, and 5.8 hrs for accuracies of 90 and 93%, respectively. Here, the learning time of *f*-BDNNN is only 10% of the learning time of the BP-based learning process without hyperparameter tuning (i.e., BP-SNN-1) for the 90% accuracy. In practice when hyperparameter modulation in required, the learning time *f*-BDNN is much lower in percentage, which can be 1% of that of the BP-SNN-3. Furthermore, the *e*-BDNN can learn two times faster than *f*-BDNN.

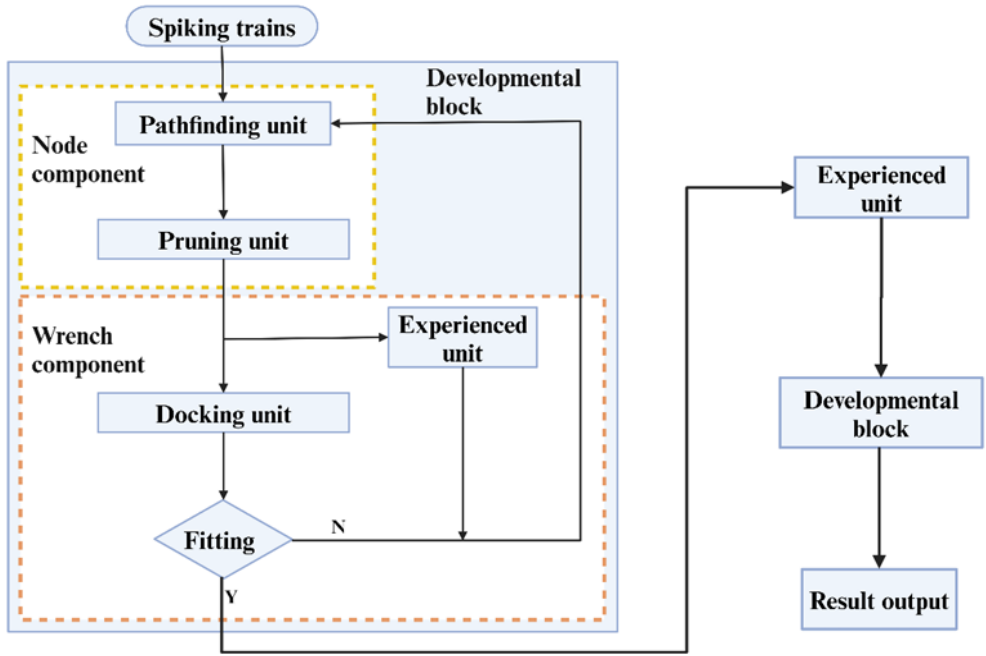

**Fig S1**. The flowchart of the learning with BDNN upon input spike trains for object classification by grasp.

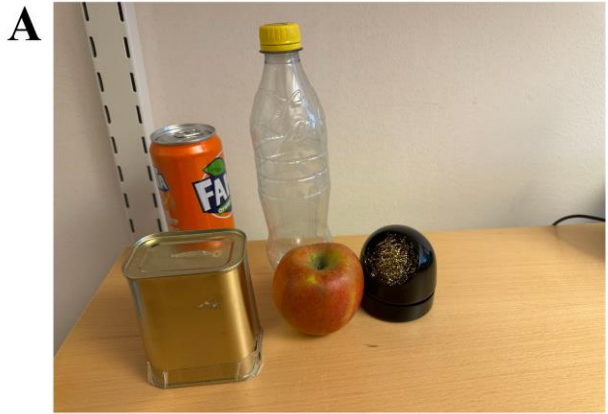 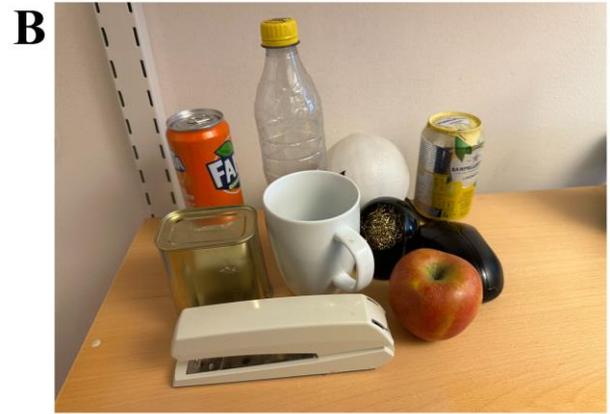
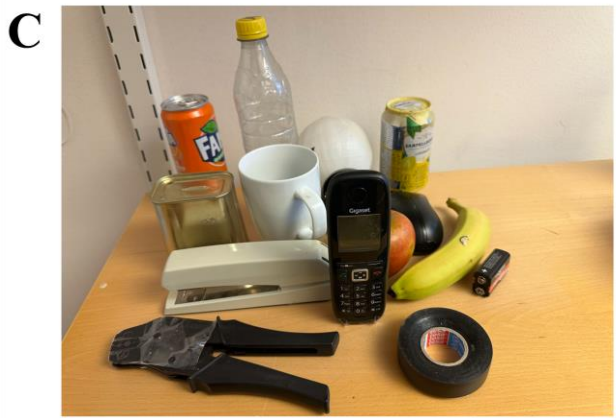 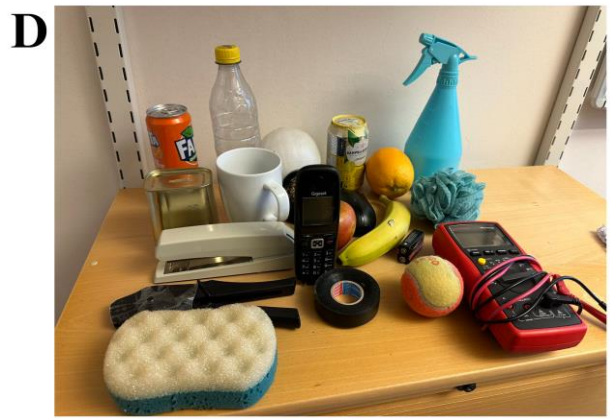

**Fig. S2**. **Photos of the objects used in grasp**. (A) Class 5-0, (B) Class 10-0, (C) Class 15-0, (D) Class 20, the pool which were used to generate tactile datasets to demonstrate the experienced learning in this work.

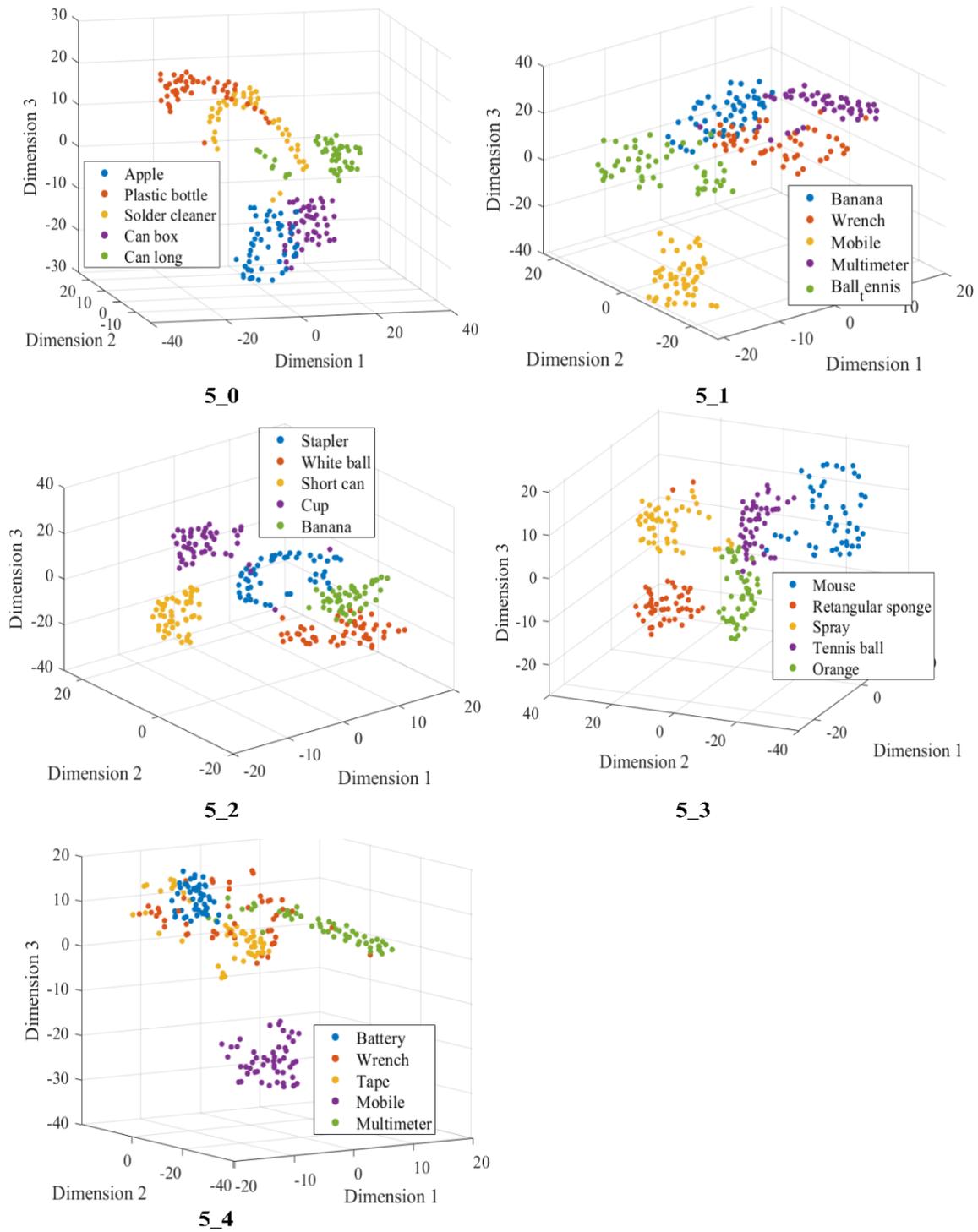

**Fig. S3. The feature distribution of different 5 objects in each class (Class 5-0 to Class 5-4) visualized by T-SNE.**

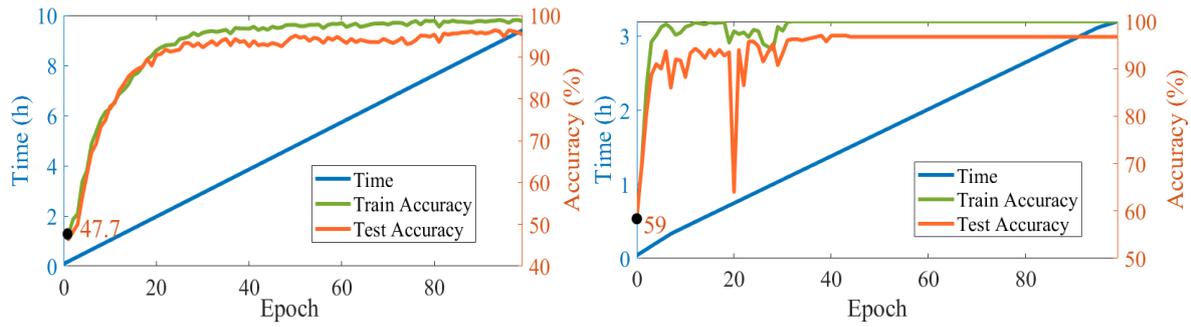

**Fig. S4. The generalization of BP-SNN (left) and BP-CSNN (right)**. 10-object class (Class 10-0) was used for the training. The 10 objects are apple, plastic bottle, solder cleaner, can box, long can, mouse, stapler, white ball, short can and cup. The initial weight is based on the pre-trained model for learning the 5-object class (Class 5-0). The 5 objects are apple, plastic bottle, solder cleaner, can box and long can. The accuracy is only 47.7%, and 59% when the BP-SNN and the BP-CSNN trained upon Class 5-0 learned Class 10-0 at the first loop (epoch), respectively.

| Datasets | No. of hidden neurons | Testing Acc. (%) |
|---|---|---|
| 5-0-0 | 63 | 100 |
| 5-0-1 | 61 | 100 |
| 5-0-2 | 67 | 99 |
| 5-1-0 | 138 | 99 |
| 5-1-1 | 108 | 97 |
| 5-1-2 | 162 | 97 |
| 5-2-0 | 100 | 98 |
| 5-2-1 | 104 | 97 |
| 5-2-2 | 108 | 98 |
| 5-3-0 | 121 | 97 |
| 5-3-1 | 158 | 98 |
| 5-3-2 | 143 | 98 |
| 5-4-0 | 150 | 94 |
| 5-4-1 | 152 | 94 |
| 5-4-2 | 168 | 95 |
| 10-0-0 | 391 | 98 |
| 10-0-1 | 359 | 97 |
| 10-0-2 | 372 | 98 |
| 10-1-0 | 424 | 93 |
| 10-1-1 | 397 | 94 |
| 10-1-2 | 378 | 94 |
| 10-2-0 | 407 | 93 |
| 10-2-1 | 345 | 90 |
| 10-2-2 | 443 | 92 |
| 10-3-0 | 380 | 96 |
| 10-3-1 | 420 | 97 |
| 10-3-2 | 270 | 96 |
| 10-4-0 | 499 | 90 |
| 10-4-1 | 393 | 90 |
| 10-4-2 | 408 | 90 |
| 15-0-0 | 336 | 92 |
| 15-0-1 | 346 | 93 |
| 15-0-2 | 427 | 95 |
| 15-1-0 | 440 | 93 |
| 15-1-1 | 415 | 91 |
| 15-1-2 | 431 | 91 |
| 15-2-0 | 456 | 94 |
| 15-2-1 | 496 | 94 |
| 15-2-2 | 432 | 93 |
| 15-3-0 | 362 | 91 |
| 15-3-1 | 551 | 93 |
| 15-3-2 | 508 | 92 |
| 15-4-0 | 407 | 91 |
| 15-4-1 | 420 | 91 |
| 15-4-2 | 554 | 93 |

**Table S 1 BDNNs constructed upon different input datasets.** Repeating the learning three times upon the same dataset, three BDNNs were generated. The label of a dataset starts the number of objects, following by the label number indicating different combination of objects in the class, ended by the label number indicating different repeated learning processes.

| Models | Accuracy (%) | Number of hidden Neurons | Training time (hrs) | Learning time (hrs) | Hyper Param. modulation |
| --- | --- | --- | --- | --- | --- |
| **$f$-BDNN-1** | 90 | 428 | **1.94** | **1.94** | **No** |
| **$f$-BDNN-2** | 92 | 725 | **9.4** | **9.4** | **No** |
| **$e$-BDNN-1** | 90 | 379 | **0.73** | **0.73** | **No** |
| **$e$-BDNN-2** | 93 | 655 | **5.8** | **5.8** | **No** |
| BP-SNN-1 | 89.4 | 50 | 29.1 (100 ep.) 19.5 (67 ep.) | 29.1 (100 ep.) | Yes |
| BP-SNN-2 | 89 | 50, 50 | 37(100 ep) | 59(100 ep) | Yes |
| BP-SNN-3 | 91.0 | 100 | 88.5 (100 ep.) 63.7 (72 ep.) | 154.6 (100 ep.) | Yes |
| BP-CSNN) | 86 | 100,128 | 13.3 (100.ep) | 13.3 (100 ep) | Yes |
| BP-CSNN | 88.2 | 32,32,128 | 12.8 | 26.1 | Yes |
| BP-CSNN | 87.6 | 50,32,128 | 15.4 | 41.5 | Yes |
| BP-CSNN | 89.5 | 32,32,32,16,64 | 11.6 | 53.1 | Yes |

**Table S2** Comparison between BDNNs and the BP-based learning procedure with SNN and CSNN models upon the dataset of the 20 objects, i.e., $X(O_{20})$. All algorithms were run on the same hardware.